\documentclass{emulateapj}

\catcode`\"=\active\let"=\"

\def\3{{\ss} }

\def\c12{{1\over 2}}

\def\plusplus{\raise 0.3ex\hbox{${\scriptstyle ++}$}{}}

\newcommand{\kms}{\,{\rm km \, s^{-1}}}
\newcommand{\oversim}[2]{\protect{\mbox{\lower0.5ex\vbox{%
   \baselineskip=0pt\lineskip=0.2ex
   \ialign{$\mathsurround=0pt #1\hfil##\hfil$\crcr#2\crcr\sim\crcr}}}}} 
\newcommand{\simgreat}{\mbox{$\,\mathrel{\mathpalette\oversim>}\,$}} 
\newcommand{\simless} {\mbox{$\,\mathrel{\mathpalette\oversim<}\,$}} 

\begin{document}

\title{No evidence for internal rotation in the remnant core of the Sagittarius dwarf} 
\shorttitle{No evidence for internal rotation in the Sgr dwarf}
\shortauthors{Pe\~{n}arrubia et al.}
\slugcomment{Submitted to ApJ}

\author{Jorge Pe\~{n}arrubia\altaffilmark{1}, Daniel B. Zucker\altaffilmark{2,3}, Mike J. Irwin\altaffilmark{1}, Elaina A. Hyde\altaffilmark{2}, Richard R. Lane\altaffilmark{4,5}, Geraint F. Lewis\altaffilmark{5}, Gerard Gilmore\altaffilmark{1}, N. Wyn Evans\altaffilmark{1}, Vasily Belokurov\altaffilmark{1}  }
\affil{$^1$Institute of Astronomy, University of Cambridge, Madingley Road, Cambridge CB3 0HA, UK}
\affil{$^2$Department of Physics and Astronomy, Macquarie University, NSW 2109, Australia} 
\affil{$^3$Australian Astronomical Observatory, PO Box 296, Epping, NSW 1710, Australia}
\affil{$^4$Departamento de Astronom\'ia, Universidad de Concepci\'on, Casilla 160 C, Concepci\'on, Chile}
\affil{$^5$School of Physics, A28, University of Sydney, NSW 2006, Australia}
\altaffiltext{1}{Email: jorpega@ast.cam.ac.uk}

\begin{abstract}
We have conducted a spectroscopic survey of the inner regions of the Sagittarius (Sgr) dwarf galaxy using the AAOmega spectrograph on the Anglo-Australian Telescope. We determine radial velocities for over 1800 Sgr star members in 6 fields that cover an area 18.84 deg$^2$, with a typical accuracy of $\approx 2\kms$. Motivated by recent numerical models of the Sgr tidal stream that predict a substantial amount of rotation in the dwarf remnant core, we compare the kinematic data against N-body models that simulate the stream progenitor as (i) a pressure-supported, mass-follows-light system, and (ii) a late-type, rotating disc galaxy embedded in an extended dark matter halo. We find that the models with little, or no intrinsic rotation clearly yield a better match to the mean line-of-sight velocity in all surveyed fields, but fail to reproduce the shape of the line-of-sight velocity distribution. This result rules out models wherein the prominent bifurcation observed in the leading tail of the Sgr stream was caused by a transfer from intrinsic angular momentum from the progenitor satellite into the tidal stream. It also implies that the trajectory of the young tidal tails has not been affected by internal rotation in the progenitor system. Our finding indicates that new, more elaborate dynamical models, in which the dark and luminous components are treated independently, are necessary for {\em simultaneously} reproducing both the internal kinematics of the Sgr dwarf and the available data for the associated tidal stream.

\end{abstract}

\section{Introduction}\label{sec:int}
Serendipitously discovered in a kinematic survey of the outer regions of the Galactic bulge, the Sagittarius (Sgr) dwarf galaxy (Ibata et al. 1994) displays one of the clearest cases of hierarchical galaxy formation in action, as it is currently being cannibalized by the Milky Way's tidal field and is close to full tidal disruption (Velazquez \& White 1995; Ibata et al.1997; Niederste-Ostholt et al. 2010). The existence of an associated stellar stream that fully wraps around the Galaxy (Mateo et al. 1996; Ibata et al. 2001; Mart\'inez-Delgado et al. 2001, 2004; Dohm-Palmer et al. 2001; Majewski et al. 2003; Belokurov et al. 2006) provides clear-cut evidence that the tidal stripping of stars commenced several orbital periods ago (Edelsohn \& Elmegreen 1997; Helmi \& White 2001; Law et al. 2005; Fellhauer et al. 2006).

However, in spite of the extensive theoretical efforts devoted to reproducing the position and velocity of the Sgr system (remnant core plus stellar stream), there currently is no model able to describe the wealth of existing observational data. A number of aspects have proven particularly difficult to understand. First, the width and velocity dispersion of the detected parts of the stream suggest that the stellar material was tidally stripped in a relatively recent epoch ($\sim 2$--3 Gyr ago). This seems at odds with the modelled orbit of the progenitor dwarf, which has an estimated perigalacticon of $\simeq 15$ kpc and a period $\simeq 1$ Gyr (Ibata et al. 1997; Zhao 1998). Jiang \& Binney (2000) showed that the relatively young age of the stream could be explained if the progenitor mass was originally as large as $\sim 10^{11} M_\odot$ and its accretion onto the Milky Way (MW) occurred at a large Galactocentric distance, $\simgreat 200$ kpc. In this scenario, dynamical friction would be responsible for dragging the Sgr dwarf to its current orbit on a time-scale that spans several orbital periods. 

Alternatively, motivated by the discovery in Sloan Digital Sky Survey data (SDSS; York et al. 2000) of a bifurcation in the leading tail of the Sgr stream (Belokurov et al. 2006), Fellhauer et al. (2006) explored numerical models where the original progenitor mass was small, $10^8$--$10^9 M_\odot$. 
In these models the role played by dynamical friction becomes inconsequential. As a result of the null orbital decay, the mass-loss process spans over a large number of orbital periods, which leads to the formation of multiple stream wraps. If the shape of the MW Galactic potential is close to spherical, these wraps may appear as bifurcated streams on the sky. Recent estimates of the stellar mass in the known pieces of the Sgr stream have lent some support to this scenario, as the coadjuted brightness of the Sgr system seems comparable to those of the M31 dwarf ellipticals (dEs) NGC 147 and NGC 185 (Niederste-Ostholt et al. 2010), whose dynamical masses are $\sim 10^{9} M_\odot$ (Geha et al. 2010). 

 However, detailed analysis of SDSS observations suggest that both arms of the bifurcated stream may exhibit similar distances, velocities and metallicity distributions (Yanny et al. 2009; Niederste-Ostholt et al. 2010), which is hard to accommodate by models that explain the bifuraction as two 
wraps of different ages, or as independent streams from different progenitors.

In a recent contribution Pe\~narrubia et al. (2010, hereafter P10) were able reproduce the existing data on the leading tail bifurcation by adopting N-body models in which the Sgr dwarf galaxy was originally a late-type, {\em{rotating}} disc galaxy embedded in an extended dark matter halo, rather than a non-rotating, pressure-supported dwarf spheroidal galaxy (dSph), as previously thought. In this scenario, the two arms are both leading the galaxy, and arise naturally during consecutive pericentric passages, which occur at different orientations of the Sgr disc relative to its orbit. Furthermore, these authors showed that internal rotation can efficiently alter the trajectory of the stream with respect to the orbit of Sgr, which might indirectly affect any constraint on the shape of the MW halo derived from pressure-supported dwarf models. 
It is thus clear that understanding the bifurcation has important implications for the
Galactic potential, which has been inferred from the Sgr stream to be variously spherical (Ibata et al. 2001), oblate (Johnston
et al. 2005), prolate (Helmi 2004), or triaxial (Law \& Majewski 2010, hereafter LM10).

Clues to the true origin of the Sgr dwarf (dSph, dE or late-type spiral galaxy), as well as to the shape of the MW dark matter halo, can be gleaned from the internal kinematics of the remnant core. Here we present the results of a spectroscopic survey of the Sgr remnant core using the AAOmega instrument at the Anglo-Australian Telescope (AAT).

\begin{figure}
\plotone{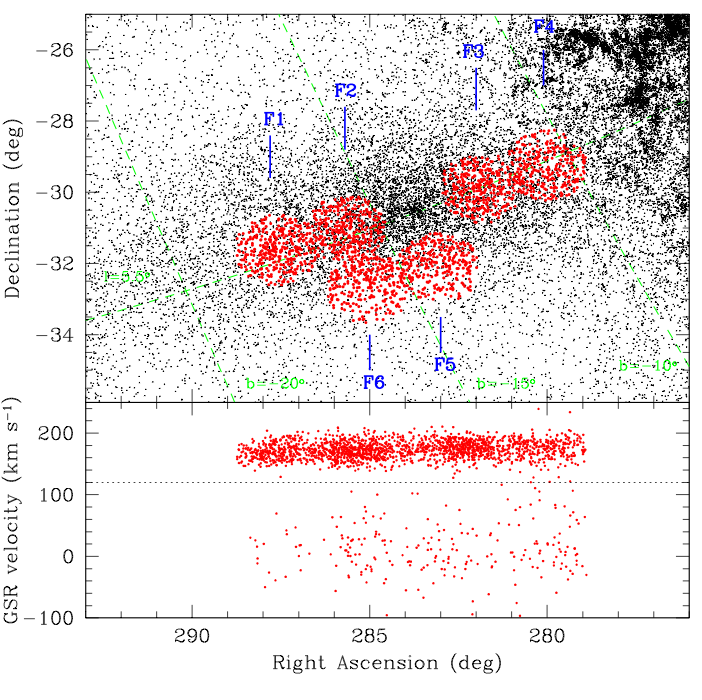}
\caption{{\it Upper panel}: the spatial distribution of selected 2MASS candidate Sgr K and M-giants (see text for details). In red (bold points) we
highlight stars with measured spectra from 6 fields (F1, ..., F6) of 2 degree diameter.
The large concentration of stars toward the upper right denotes the main
Milky Way contamination.  {\it Lower panel}: the GSR velocity distribution
of the candidates observed.  The dashed line at $120 \kms$ is our selection
boundary for Sgr members and provides a clean separation from Milky Way
foreground interlopers.}
\label{fig:fields}
\end{figure}

\section{Observations}\label{sec:data}
We observed candidate Sgr red giant stars with the AAOmega fiber-fed spectrograph on the AAT the nights of 2010 June 3 -- 6. AAOmega obtains spectra of $\sim400$ targets simultaneously over a 2-degree field of view, employing a dichroic to send light to separate blue and red spectrographs. We used 1500V ($R \sim 3700$) and 1700D ($R \sim 10000$) gratings in the blue and red, to cover the regions around Mg I 5170\AA\ and the Ca II 8500\AA\ triplet, respectively. Although most of the observing run was lost to clouds, we were able to obtain high-quality data on 6 out of 9 intended fields in the central regions of the Sgr dwarf. The data were reduced using the Australian Astronomical Observatory's {\tt 2dfdr} software package.

Target stars were selected from the 2MASS PSC catalogue (Skrutskie et al. 2006). The photometry was first reddening-corrected using the Schlegel et al. (1998) extinction maps, and then candidates were required to satisfy the criteria $9 < K_{\rm s} < 13$ and $J - K_{\rm s} > 2.222 - 0.111K_{\rm s}$, i.e., likely Sgr K and M-giants. The location on the sky of these candidate Sgr stars is shown in the upper panel of Fig. 1; objects for which we obtained spectra are shown using bold
points. 
The lower panel of Fig. 1 shows the measured Galactic Standard of
Rest (GSR) velocity distribution as a function of RA.  Velocities
were derived from cross-correlating the Ca II triplet region of the
spectrum (8450\AA -- 8750\AA) with a model template spectrum (see
Battalgia et al. 2008 for further details), and were corrected to a GSR frame.Star-by-star comparison of these results with an independent cross-correlation of the spectra, using the
IRAF task {\tt fxcor} and a synthetic Ca II triplet spectrum, yielded mean offsets of $\simless 0.5\kms$,
with a typical standard deviation of $< 1.5 \kms$.

MW foreground contamination is readily recognized
from the spatial density variations visible in the upper panel coupled
with the clean separation in a GSR frame from Sgr members ($v_{\rm GSR} >
120 \kms$) that appears in the lower panel.  We find that 1805 stars
from our spectroscopic sample satisfy the velocity cut and are considered
as Sgr dwarf members in this paper.

\section{Theoretical models}\label{sec:models}
 To analyze whether the Sgr dwarf possesses internal angular momentum we compare the above kinematic data against two types of N-body models: one in which the Sgr dwarf galaxy is composed of a late-type exponential disc embedded in a non-singular isothermal dark matter halo ({\it rotating model}), and another where the stellar and the dark matter particles follow a spherical King (1966) model ({\it pressure-supported model}). The main characteristics of the rotating model are the following (for details see P10): the initial mass is $M_0= m_d+m_h = 3.5\times 10^8M_\odot + 2.4\times 10^9M_\odot \approx 2.8\times 10^9M_\odot$, where $m_d$ and $m_h$ denote the disc and dark matter halo masses. The initial disc scale-length and the halo core radius are $R_d\simeq 0.9$ kpc and $r_h=0.5 R_d$, respectively. The rotation curve of this model peaks at $\approx 43\kms$. 
For the pressure-supported model we adopt similar parameters as those derived by LM10 in order to reproduce most of the existing observational constraints on the Sgr stream, i.e. an initial mass of $M_0=6.0\times 10^8 M_\odot$, a core radius $R_K=0.85$ kpc, and a King tidal radius $R_t=5.02$ kpc.

We assume a Galactic potential composed of disc, bulge and dark matter halo (see P10 for details). A brief description of this potential is the following: the disc component follows a
Miyamoto-Nagai (1975) model with a mass $M_d=7.5\times 10^{10}
M_\odot$, and radial and vertical scale lengths $a=3.5$ kpc and
$b=0.3$ kpc. The MW bulge follows a Hernquist (1990) profile with a
mass $M_b=1.3\times 10^{10}M_\odot$ and a scale radius $c=1.2$ kpc.
The MW dark matter halo is modelled as a Navarro, Frenk \& White (1997) profile with a virial
mass $M_{\rm vir}=10^{12}M_\odot$, virial radius $r_{\rm vir}=258$ kpc
and concentration $c_{\rm vir}=12$ (Klypin et al. 2002). In order to construct a Sgr dwarf model that matches the position and velocities of the observed stream pieces, we follow LM10 in assuming that the Galactic halo is triaxial in shape. This is done by introducing
elliptical coordinates with the substitution $r\rightarrow m$, where
$m^2={x^2}/{a^2}+{y^2}/{b^2}+{z^2}/{c^2}$, and $(a,b,c)$ are
dimensionless quantities with values $(a,b,c)\simeq (0.61,1.34,1.22)$.

We use test particles to
integrate the orbit of the Sgr dwarf back in time from the current position and velocity of the remnant core in order to derive initial conditions for our N-body models. For simplicity, we hold the halo parameters fixed through the evolution of our Sgr models, given that tidal streams are barely sensitive to the past evolution of the host potential
(Pe\~narrubia et al. 2006). We use an total integration time of 2.5 Gyr (approximately 2.5 orbital periods), as this is the (minimum) time required to reproduce the observed pieces of the Sgr stream (see P10).
The current position of the Sgr dwarf in galactocentric coordinates is $(D,l,b)_{\rm Sgr}=(25 {\rm kpc}, 5^\circ.6, -14^\circ.2)$. The velocity vector can be derived from the line-of-sight velocity $v_{\rm
  los}=137$ km s$^{-1}$  (Ibata et al. 1997), and the proper motion measurements of Dinescu et al. (2005), who find $(\mu_l\cos
b,\mu_b)=(-2.35\pm 0.20, -2.07 \pm 0.20)$ mas/yr. Adopting standard values for the Galactocentric position of the sun $(X,Y,W)_\odot=(-8,0,0)$ kpc, the Local Standard of Rest and the solar peculiar motion $(U,V,W)_\odot=(10,225,7)\kms$ (Binney \& Merrifield 1998), we find a Galactocentric space motion
is $(\Pi,\Theta,W)_{\rm Sgr}\simeq (110,-37,264)$ km s$^{-1}$. 
After integrating our N-body representations of the Sgr stream progenitor forwards in time for 2.5 Gyr, the rotating and pressure-supported models have respectively lost 42\% and 48\% of their initial stellar masses to tides.

Fig.~2 shows the stellar remnants of our rotating (upper panel) and pressure-supported (lower panel) dwarf galaxy models. N-body particles are color-coded according to their mean line-of-sight velocity in the GSR frame. This figure illustrates a few interesting points. First, both dwarf models show velocity gradients throughout the dwarf remnant core. Indeed, this is a feature expected in models where the Sgr dwarf is assumed to have originally been a late-type galaxy. In particular, the upper panel of Fig.~2 shows that, although tidal mass stripping removes a large fraction of the original angular momentum in the progenitor Sgr disc, the remnant core still rotates with a velocity amplitude $\sim 20$ km s$^{-1}$, which translates into a net velocity difference of $\sim 40$ km s$^{-1}$ between stars at declinations above and below $\delta \sim -30^\circ$. 

The presence of a velocity gradient in the pressure-supported model, however, may be a result of the line-of-sight projection of the orbital velocity vector, rather than a manifestation of internal rotation. Since the orbital trajectory (dashed line) is practically perpendicular to the MW disc (dotted line), we expect a velocity gradient to principally manifest at different latitudinal directions with an amplitude $\Delta v_{\rm app} = \Delta b (d {\bf \hat r }/db \cdot {\bf v})\simeq -231 \Delta b \kms$, where $\Delta b=b-b_{\rm Sgr}$. Given that the maximum separation of our fields (white circles) is $|\Delta b|\simeq 4^\circ$, we expect the line-of-sight velocity to increase/decrease approximately 16 $\kms$ with respect to the systemic velocity of the dwarf in the fields that lie closest to/furthest from the MW plane. 

The second interesting point refers to the direction of the velocity gradient. Whereas the rotating model predicts an increasing velocity towards higher Right Ascensions, the pressure-supported model shows the opposite trend. Given the accuracy of our velocity measurements (typically $2 \kms)$ and the large number of Sgr members in our sample ($N\approx 1800$), the above model predictions are straightforward to test.

\begin{figure}
\plotone{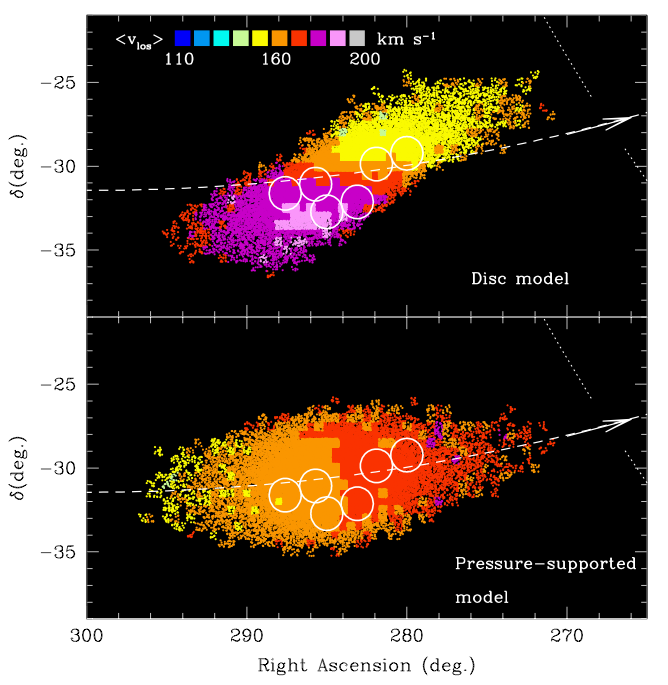}
\caption{The remnant cores of the Sgr dwarf rotating ({\it upper panel}) and pressure-supported ({\it lower panel}) models. Particles are color-coded according to the mean line-of-sight velocity. Circles denote the location of the 2dF camera fields. Dashed lines and arrows mark the orbital trajectory of the Sgr dwarf. The location of the Galactic equator is shown with a dotted line for ease of reference. Note that the remnant core of a pressure-supported model shows a velocity gradient with an amplitude $\sim 10$ km/s }
\label{fig:core}
\end{figure}

\section{Models vs. data}\label{sec:comp}
\begin{figure}
\plotone{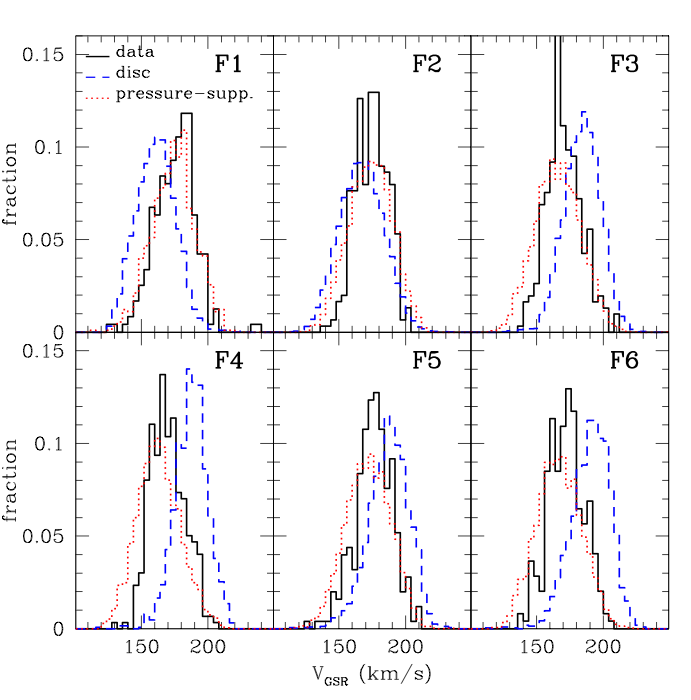}
\caption{Line-of-sight velocity (in Galactic Standard of Rest coordinates, see text) histograms in our 6 AAO fields (solid lines) in bins of $\Delta v=4$ km/s (note that the Sgr member stars have velocity measurements with uncertainties $\epsilon_v<\Delta v$). For ease of comparison we show the predictions from pressure-supported (dotted lines) and rotating (dashed lines) models. Note that the pressure-supported models provide a better fit to the data. }
\label{fig:vdist}
\end{figure}

Fig.~\ref{fig:vdist} shows the comparison of the velocity distribution expected for rotating (dashed lines) and pressure-supported (dotted lines) dwarf models in each of the fields where data (solid lines) exists. A number of revealing points can be gleaned from this figure. The first is that we observe a clear variation in the velocity peak along the major axis of the dwarf. For example, comparing fields F1 and F4, which  are separated by 8 degrees (see Fig.~1), the velocity distribution peaks at $\simeq 185 \kms$ and $\simeq 165 \kms$, respectively. The amplitude, as well as the sense, of the velocity shift is consistent with the variation in the mean line-of-sight velocity expected from a projection effect, as previously discussed.  Further evidence for this interpretation can be found in the absence of a mean velocity gradient along the minor axis, as the peak velocity in both the F2 and F6 fields is $175 \kms$.

Indeed, it is clear that the pressure-supported model provides a better match to the mean velocities measured throughout the Sgr remnant core than the rotating model, which predicts a velocity trend that is nearly opposite to the one observed. Our data thus seem to rule out the scenario proposed by P10, wherein the bifurcation observed in the Sgr stream  would be caused by a transfer of internal angular momentum from the progenitor satellite into the stream.

However, it is worth noting that, although the pressure-supported model
provides a reasonable qualitative description of the mean velocities observed throughout the remnant core, the shape of the line-of-sight velocity distribution appears considerably more leptokurtic than predicted in all the surveyed fields, except perhaps in F1. This result is hardly surprising, considering that the dwarf model adopted by LM10 in order to reproduce the properties of the Sgr stream assumes for simplicity that the stellar and the dark matter particles follow the same density profile. Our data clearly indicate that better kinematic coverage of the Sgr core plus more sophisticated mass modelling should provide strong constraints on the original distribution of stars and dark matter in this galaxy.

\section{Discussion}

Despite the prodigious theoretical efforts undertaken to reproduce the
characteristics of the Sagittarius tidal stream (e.g. Ibata et
al. 2001; Mart{\'{\i}}nez-Delgado et al. 2004; Helmi 2004; Law et
al. 2005, 2010; Fellhauer et al. 2006), two aspects of the stream have proven particularly challenging to understand.  First, the leading tail of the Sgr stream is bifurcated,
with both arms exhibiting similar distances, velocities and
metallicity distributions (Yanny et al. 2009; Niederste-Ostholt et al. 2010), 
which would appear to refute earlier models that explain the bifurcation as two
wraps of different ages (Fellhauer et al. 2006), or as independent
streams from different progenitors. Second, the position on the sky and the heliocentric velocities of stream members suggest the possibility of the MW dark matter halo being triaxial in shape (LM10). However, the orientation of the favored triaxial halo model is puzzling, for its intermediate axis appears to be aligned with the spin vector of the MW
disc. This is hard to understand, as circular orbits about the
intermediate axis are unstable, raising questions about the formation
of the Galactic disc.

In a recent contribution, P10 showed that internal rotation in the progenitor dwarf might help to explain the origin of the bifurcation
in the leading tail of the Sgr stream, as well as the puzzling shape and orientation of the MW dark matter halo derived by LM10. In particular, P10 found that the same mechanism that gave rise to a bifurcated leading tail, i.e. a transfer of internal angular momentum into the tidal stream, could have also altered the orientation of the leading tail with respect the progenitor's orbit, directly affecting the constraints on the halo shape derived from stream models that assume a pressure-supported progenitor galaxy. To check this scenario, these authors proposed a kinematic survey of the Sgr remnant core, which according to their models should exhibit a clear signature of internal rotation.

Here we have conducted a spectroscopic survey of the inner regions of the Sgr dwarf using the AAOmega instrument on the AAT. Our survey covers 6 fields with a total area of $6\pi \simeq 18.8$ deg$^2$ and provides radial velocities for more than 1800 Sgr dwarf member stars with a typical accuracy of $\approx 2$ km/s. 
We have compared our velocity measurements against numerical N-body simulations that adopt rotating (P10) and pressure-supported (LM10) dwarf N-body models. We find that models with little or no intrinsic rotation provide a better match to the mean line-of-sight velocity  in each of the surveyed fields, hence refuting the scenario proposed by P10, but fail to reproduce the observed line-of-sight velocity distribution, which appears considerably more leptokurtic (or peaked), especially in the inner-most fields.

Interestingly, the absence of intrinsic rotation may be consistent with recent theoretical models that propose 
tidal stirring as the driving mechanism for the formation of the Sgr dwarf (Lokas et al. 2010). In this scenario discs are transformed into spheroids through a complex process that involves tidal mass loss, bar formation/disruption and ram stripping. However, although these models are able to reproduce the elongated shape of the Sgr dwarf, as well as the absence of intrinsic angular momentum in the remnant core, they cannot explain the presence of a bifurcated tidal tail with the observed characteristics.

Further theoretical work is clearly necessary in order to reproduce the internal kinematics of the remnant core, as well as the locations and velocities of the associated stream, {\em simultaneously}. The kinematic data presented in this contribution will provide useful constraints for future numerical models that treat the dark and baryonic material as separate dynamical components (e.g. Lokas et al. 2010). Meanwhile, the origin of the prominent bifurcation in the distribution of tidal debris of the Sgr dwarf stands as an open -- and enigmatic -- question.

\acknowledgments
JP acknowledges financial support from the Science and
Technology Facilities Council of the United Kingdom. RRL acknowledges support from the Chilean Center for
Astrophysics, FONDAP Nr. 15010003, and from the BASAL Centro de
Astrofisica y Tecnologias Afines (CATA) PFB-06/2007. GFL is grateful to the IoA for sabbatical support
 through the award of the Raymond and Beverley Sackler
 Distinguished Visitorship.

{}

\end{document}